\documentstyle{mn}

\def\today{\ifcase\month\or
 January\or February\or March\or April\or May\or June\or
 July\or August\or September\or October\or November\or
 December\fi\space\number\day, \number\year}

\def\todmy{\number\day\space\ifcase\month\or
 January\or February\or March\or April\or May\or June\or
 July\or August\or September\or October\or November\or
 December\fi\space\number\year}

\newcommand{\bdisp} {\begin{displaymath}}
\newcommand{\edisp} {\end{displaymath}}
\newcommand{\beqn} {\begin{equation}}
\newcommand{\eeqn} {\end{equation}}
\newcommand{\beqr} {\begin{array}}
\newcommand{\eeqr} {\end{array}}

\newcommand{\tal}{\it et al. \rm}

\title{Performance and accuracy of a GRAPE-3 system for
collisionless N-body simulations}
\author[E. Athanassoula, A. Bosma, J.-C. Lambert, J. Makino]
       {E. Athanassoula,$^1$, A. Bosma,$^1$ J.-C. Lambert$^1$, 
	and J. Makino,$^2$\\ 
       $^1$Observatoire de Marseille,
       2 Place Le Verrier,
       F-13248 Marseille Cedex 4, France\\
       $^2$Department of Information Science and Graphics,
College of Arts and Sciences, The University of Tokyo,
Tokyo 153, Japan}

\date{Accepted .
      Received ;
      }

\pagerange{\pageref{firstpage}--\pageref{lastpage}}
\pubyear{1997}

\begin{document}

\maketitle

\label{firstpage} 

\begin{abstract}
The performance and accuracy of a GRAPE-3 system
for collisionless N-body simulations is discussed.
After an initial description of the hardware configurations 
available to us at Marseille, and
the usefulness of on-line analysis, we concentrate on the
actual performance and accuracy of the direct summation and
tree code software. For the former we discuss the sources of round-off
errors. The standard Barnes-Hut tree code can not be used as such on a
GRAPE-3 system. Instead particles are divided into blocks and the tree
traversal is performed for the whole block, instead of for each
particle in the block separately. Then the forces are calculated by
direct summation over the whole interaction list. The performance of
the tree code depends on the number of particles in the block,  
the optimum number depending on the speed of the front end and the
number of boards. We find that the code scales as ${\cal O}(N)$ and
explain this behaviour. The time per step decreases as the tolerance 
increases,
but the dependence is much weaker than for the standard tree
code. Finally we find that, contrary to what is expected for the
standard version, the speed of our tree code increases with the
clustering of the 
configuration.  We discuss the effect of the front end and compare the
performance of direct summation and tree code on GRAPE-3 with that of
other software on general purpose computers. 

The
accuracy of both direct summation and the tree code is discussed as 
function of number of particles and softening. For this we consider
the accuracy of the force calculation as well as the energy
conservation during a simulation. Because of the increased role of the
direct summation in the force calculation, our tree code is much more
accurate than the standard one. Finally we follow the evolution of an
isolated barred galaxy using different hardware and software
in order to assess the reliability and reproduceability of
our results. We find excellent agreement between the pattern speed of
the bar in direct summation simulations run on the high precision
GRAPE-4 machines and that in direct summation simulations run on our 
GRAPE-3 system. The agreement with the tree code
is also very good provided the tolerance values are smaller than 
about 1.0. 

We conclude that GRAPE-3 systems are well suited for 
collisionless simulations and in particular those of galaxies. This is
due to their good accuracy and their high speed which allows the use
of a large number of particles.

\end{abstract}


\section{Introduction }
\label{sec:intro}
\indent

N-body simulations have come a long way since the pioneering work 
of Holmberg (1941). The ever-increasing computing power available has
allowed them to become very useful tools for the understanding of
the formation, dynamics and evolution of galaxies, and, particularly
if used in concert with analytical work and confrontation to
observations, can trigger important progress in the field. For this
reason many different codes have been written to date, each aiming for
a better accuracy and/or performance. Recent reviews of this quest
have been given by Sellwood (1987) and Athanassoula (1993). The
simplest approach, direct summation, is the one containing the least
approximations. In this case the force on a given particle is obtained
simply by adding the contributions from all other particles in the
configuration. The big disadvantage of this method is that it is
prohibitively expensive in CPU time, since the computational cost is
proportional to $N(N-1)$, where $N$ the number of particles in the
configuration. For this reason it could not be used up till now
with a number of
particles sufficient for most applications. This situation changed
drastically with the advent of the GRAPE boards, which are
special-purpose boards on which are calculated the gravitational
forces between particles. These are linked to a front end machine,
performing all the remaining tasks in the simulations. In this way
very high performance can be achieved. A good description of the
GRAPE project and the technical description of the boards can be found
in a number of papers written by the team that conceived them
(e.g. Sugimoto \tal 1990, Ebisuzaki \tal 1993, Makino \tal 1997)

In this paper we evaluate the performance and accuracy of
the Marseille GRAPE-3 systems, and compare them with those of
general purpose computers. The outline of the
paper is as follows: We give the
description of the configurations in section \ref{sec:description}. 
We are using two
types of software, direct summation (sec. \ref{sec:N2soft}) and the
tree code (sec. \ref{sec:treesoft}). We discuss in detail the performance
of the latter in section \ref{sec:tree-time} and in general the
timing of our GRAPE-3AF system in section \ref{sec:timing}. We next
discuss the accuracy of the direct summation (sec. \ref{sec:n2-accu})
and of the tree code (sec. \ref{sec:tree-accu}). Section \ref{sec:bar}
is devoted to comparisons of results obtained with these two methods and
of results obtained on GRAPE-4 hardware. The example chosen 
is the long term evolution of a barred galaxy.

\section{Description of the Marseille GRAPE systems }
\label{sec:description}
\indent

\begin{figure}
\includegraphics{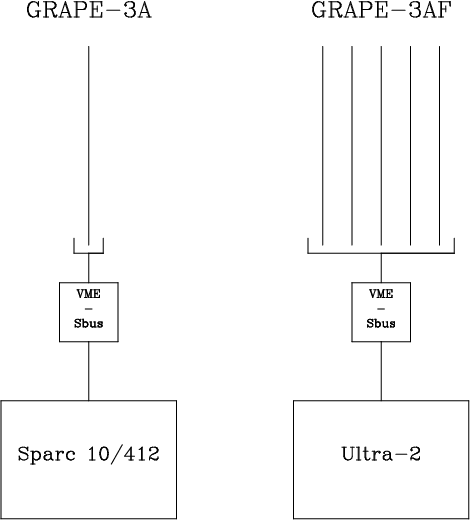}
\vspace{7.5cm}
\caption{
Schematic layout of our two GRAPE-3 systems, on the left GRAPE-3A
and on the right GRAPE-3AF.}
\label{grapes}
\end{figure}

Our group acquired a hand-wired GRAPE-3A card in March 1993. This consists of
4 LSI chips operating at 20 MHz and gives a peak speed equivalent of
more than 2 Gflops. It is coupled via a Solflower
SFVME-110 Sbus/VMEbus converter to a Sparc 10/412 workstation, which
drives it.  The configuration is
illustrated in the left panel of Figure~\ref{grapes}. This hardware is
limited to 32~768
particles. It is, however, possible to use more particles than the hardware 
limit simply by dividing the total number of particles into packages of 32~768 
or less, presenting one package at a time to the board, and then adding the 
contributions of all packages on the front end machine. Further 
information on the GRAPE-3A chips and the relevant software can be found in 
Ebisuzaki {\it et al.} (1993) and Makino \& Funato (1993).

In August 1994 we acquired five GRAPE-3AF boards, each having 8 
chips, giving us a peak speed equivalent of more than 20 Gflops. 
These are printed-circuit boards and occupy the five slots of 
a Solflower SFVME-110 Sbus/VMEbus
converter, which links them to the host machine. As such we have used 
consecutively a Sparc 10/41, a Sparc 10/512, an Ultra 1/170 and, since
August 1996, an Ultra 2/200 with two processors. The usefulness of the
second processor will be
discussed later in this section. The configuration is illustrated 
in the right panel of
Figure~\ref{grapes}. GRAPE-3AF cards are hardware limited to 131 072 
particles, yet it is possible, as for the GRAPE-3A boards, to use 
them for a larger number of particles with the help of 
appropriate software. 

Every GRAPE-3A or GRAPE-3AF board has one memory unit. In case of multiboard 
systems, as for our GRAPE-3AF system, the host computer sees the system as one 
unit, since the memory units share the same address space. Thus when the host 
computer sends data to the memory, these are stored identically on the memory 
units of all boards. Different boards calculate forces on different particles, 
so that if a system has $N_{chips}$ chips the forces will be calculated 
simultaneously for $N_{chips}$ particles.
Because of the nature of the N-body problem it is easy to use all chips in 
parallel, since the 
forces on different particles can be calculated independently. 

Considerable gain in time can be achieved if the data transmission
overlaps partly with the force calculation on the chips (Okumura \tal
1993). Thus when the data has been loaded to the first board this board can
start calculating, without waiting for the data to be loaded on the
second board etc. This amounts to a significant saving in the case of
five boards, as in the Marseille system. 

Since GRAPE-3 boards are meant to be used only for collisionless
simulations they 
use low accuracy arithmetic. 14 bits are used to represent the masses,
20 bits for 
the positions and 56 bits for the forces. As will be argued below, the 
accuracy thus obtained is sufficient for most collisionless
simulations. A more extensive discussion of the round-off errors is
given in section \ref{sec:soft}.

The output from our simulations consists of masses, positions and
velocities of all particles, as well as in some cases, potentials. For
real*4 accuracy this means 32 bytes per particle and time step saved. 
To keep the necessary disc space to manageable limits we have recourse
to on-line analysis, in which we anticipate which physical quantities
need to be extracted from the data. Storing these can be done relatively
frequently, so that the analysis can be done with adequate time
resolution. Such analysis can include the amplitude and phase of
several Fourier components of the density of individual galaxies,
their energy and angular momentum, or that of their subcomponents, or
whether individual galaxies in a cluster or group have merged or not.
Furthermore bitmaps containing frames of the particle
positions at a given time step can be stored to construct short
``movies'', which are helpful in revealing features of interest.

We found that a most suitable way to handle the load of the on-line
analysis is to use a second processor on the same workstation. The first
processor running the GRAPE boards then spawns at regular intervals a
task starting analysis scripts. Certain tasks, such as calculating the
amount of mass still bound to a companion or to a given galaxy in a
group, involve  
potentials or forces 
and are calculated faster on a GRAPE
board.
In such cases we have recourse to the computer which runs the GRAPE-3A
board.

\section{Software }
\label{sec:soft}

\subsection{The direct method}
\label{sec:N2soft}
\indent

The direct method on GRAPE-3 is quite straightforward.  The GRAPE
hardware calculates the pairwise interactions and sums the forces on
each particle from all other particles in the system. More detailed
information, together with some examples, has been given by Makino \&
Funato (1993). Here we will discuss how the round-off
errors are generated in GRAPE-3. 

The round-off error in the GRAPE-3 hardware has two main
origins (Makino, Ito \& Ebisuzaki 1990). The first one is the round
off error generated when the positions are converted from
floating-point number
format to fixed-point format. GRAPE-3
performs the subtraction ${\bf x}_i - {\bf x}_j$ in 20-bit
fixed-point format, and  a resolution of 1/1024 is typically used, where
the size of the system is taken to be of order unity. For the pairwise
force, the relative round-off error due to this is
${\cal O}(10^{-3}/r_{ij})$. In other words, the error is larger for nearby
pairs than for far-away ones. 

The second source of round-off error is the calculation of the force
from the relative position vector. This part of the calculation is
implemented in logarithmic format with an accuracy of around 1\%, and
has a r.m.s. error of 2\% (Okumura et al. 1993). 
This error does not depend on $r_{ij}$.

In conventional 
general-purpose computers additional round-off error is
generated in the summation, since one typically adds small numbers 
(the pairwise forces) to a large number (the calculated total force). 
However, since GRAPE-3 uses a 56-bit fixed point format for the
summation no additional round-off errors are generated here.

\subsection{The tree code}
\label{sec:treesoft}
\indent 

The version of tree code we use is essentially the
same as the vectorisation scheme described by Barnes (1990). The main
difference from the standard algorithm (Barnes \& Hut 1986) is that
the particles are first divided into blocks and then the tree 
traversal is performed for a block of particles, instead of
for each particle in the system. Then GRAPE hardware is used to
calculate the forces from the nodes in the interaction list created by
this tree traversal to all particles in the block. Note that this
procedure is essentially the same as the construction of the ``local
essential tree" in the distributed-memory parallel version of the tree code
developed by the Caltech group (Salmon \& Warren 1994). However, the way
we use the obtained data is quite different. In the case of the
parallel tree code, the force on each particle is still calculated by
traversing the tree, while in the case of the GRAPE tree code, we calculate
the force by direct summation over the whole interaction list.

The speed of the tree traversal, which is performed on the front end, 
depends on the
number of particles in the block. Generally speaking, if we make the
number of particles in a block larger, we can reduce the amount of
work of the front end, but we increase the cost of the calculation on
GRAPE, since the average length of the interaction list becomes longer
(Makino 1991).  Thus, the number of particles in a block should be chosen so
that the total calculation cost is minimum. This optimal number
depends mainly on the relative speed of GRAPE and the front end, but
also, though to a smaller extent, on the number of particles 
and their distribution in the system.

It is intuitively expected that the block size should also influence 
the accuracy, since the interaction list of a given block will contain 
at least as many members as the interaction list of any particle in the
list and probably quite a bit more. This was tested by 
Barnes (1990), for values of $n_{crit}$ less than 256, and will be
further examined in section \ref{sec:tree-mase} for values of
$n_{crit}$ of a few thousand, which will be shown in section
\ref{sec:ncrit-time} to be optimum for our GRAPE configuration.

A further difference with the Barnes-Hut (1986) standard algorithm is
that instead of the conventional multipole acceptability
criterion (MAC) our code uses the minimum distance MAC (Salmon \&
Warren 1994). In this criterion
instead of the distance between a body and the center of mass of a
cell one uses the minimum distance from the body to any point in the
cell. This has the useful aspect that it is completely independent of
the contents of the cell and of the center of the mass of the bodies
in it. A further advantage of this criterion according to Salmon \&
Warren (1994) is that it does not admit the rare, yet not impossible,
unbound error they found for the conventional MAC.

Since GRAPE-3 can only calculate the forces and potentials between
pairs of particles, the tree code has only the monopole term and
neglects quadrupoles and higher order terms. However, as we shall see
in sections \ref{sec:tree-accu} and \ref{sec:bar}, 
this does not prohibit it from reaching a
satisfactory accuracy level, as one can, without much
additional cost, decrease the tolerance or increase the number of
particles. 
 
\section{Performance of the tree code }
\label{sec:tree-time}
\indent

\subsection{Dependence on the number of particles in a block }
\label{sec:ncrit-time}
\indent

In this section we examine more closely how the number
of particles in a block influences the performance of the tree code. In order
to divide the particles into blocks we descend the tree and regard a
cell as being a group if the number of particles it contains is
less than a given number, $n_{crit}$, while its parent cell contains
more particles than $n_{crit}$. Thus the number of particles in a
block can vary, but must necessarily be smaller than
$n_{crit}$. Makino (1991) showed that for a Plummer sphere\footnote{
The density of the Plummer sphere is

${\displaystyle \rho (r) = \frac {3M}{4\pi b^3} (1+\frac {r^2}{b^2})^{-5/2}}$ \\
\noindent where $M$ and $b$ are its mass and scale length}
and for
values of $n_{crit}$ roughly in the range $10 <
n_{crit} < N/100$ (where $N$ the total number of particles) the 
number of particles per block is roughly $n_{crit}/4$. 

As already mentioned in the preceding section, the performance of the
tree code will depend on the number of particles in the block and
therefore on $n_{crit}$. We also
expect it to depend on the total number of particles and the
tolerance. In order to have some quantitative estimates of how these
three parameters influence the performance we have measured the time
necessary for one time step for several values of $n_{crit}$
as well as of the tolerance and of the total number of
particles. These times were obtained by averaging over 20 time steps
in order to decrease the effect of statistical fluctuations. Plotting these
times as a function of $n_{crit}$ we find a minimum, as expected, but
it is very shallow and thus ill defined. For $N = 500~000$ we
find that the optimum range for $n_{crit}$ is between 5~000 and 
25~000, and this range shrinks towards smaller values as $N$ decreases,
so that e.g. for $N=150~000$ it is between  3~000 and 15~000. For
a large number of particles the value of the tolerance does not seem
to influence the optimum value of $n_{crit}$. For $N$ around
100~000, however, larger values of the tolerance correspond to smaller optimum
values of $n_{crit}$. The fact that the minimum is so shallow leaves
a lot of freedom in the choice of the optimum $n_{crit}$. The
trade-off in this case is between memory requirements and accuracy,
since for large $n_{crit}$ the results are more accurate, but
necessitate more memory, than for low $n_{crit}$. 

Seen the above, we can conclude that a reasonable compromise for our
GRAPE-3AF system is for $n_{crit}$ around say 7~000 or 8~000.

\begin{figure}
\includegraphics{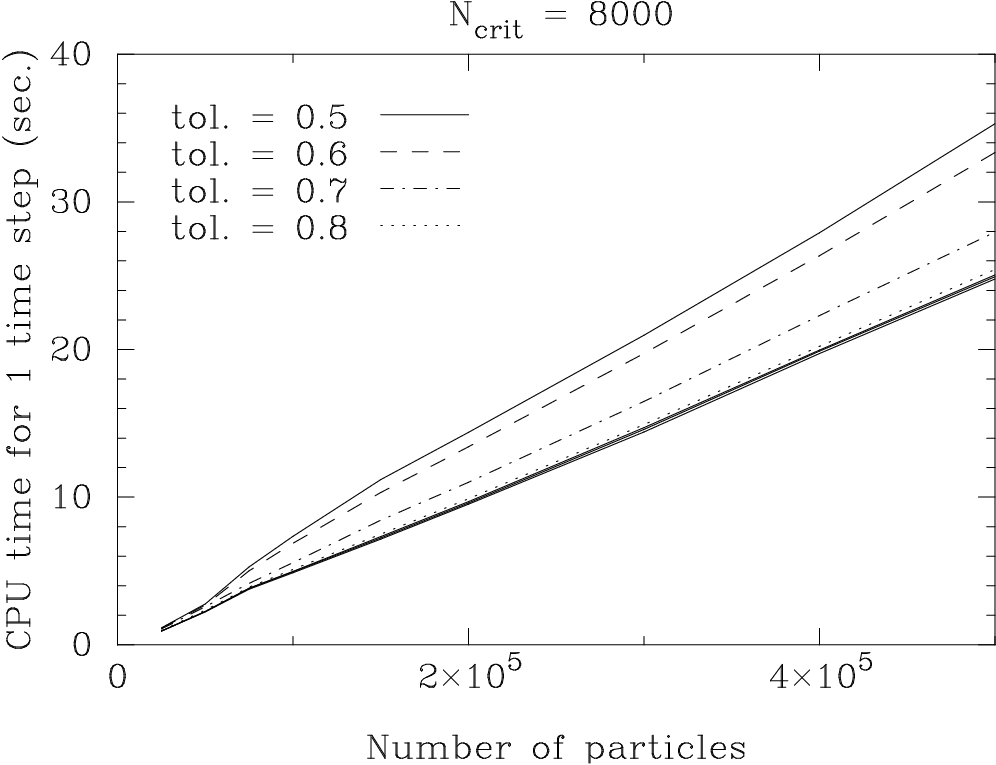}
\vspace{6.5cm}
\caption{
CPU time necessary for one time step with the tree code, as a function 
of the number of
particles in the system. The results correspond to $n_{crit} = 8~000$
and various values of the tolerance. The thick line corresponds to
values for the tolerance of 0.9, 1.0, and 1.1, which all cluster together.
}
\label{time_n}
\end{figure}

\subsection{Dependence on the total number of particles }
\label{sec:nbody-time}
\indent

Figure ~\ref{time_n} shows the CPU time per time step as a function of the
total number of particles in the system, $N$, and for various values of the 
tolerance. As expected, this time increases with $N$, but what is
interesting to note is that this increase is {\it linear}. For the
standard tree code Barnes and Hut (1986) argued that it should have an
$Nlog(N)$ dependence, while Hernquist (1987) showed that for his
vectorised implementation the CPU time was proportional to
$aNlog(N)+bN$ for a tolerance $\theta\geq 0.4$. In order to understand the time
dependence of our tree code we use the analysis of Makino (1991) which
shows that the time for the force calculation should be proportional
to $a(n_{crit},\theta)N+100N log_{10}(N\theta^{3}/23)/\theta^{3}$, where
$a(n_{crit},\theta)$ a function of $n_{crit}$ and $\theta$, given by
eq. (7) of Makino (1991). For the values of $n_{crit}$, $N$ and
$\theta$ used in our cases the first terms is much bigger than the
second one, so that the time dependence should be proportional to
$N$, in good agreement with the results of Fig. \ref{time_n}.

\subsection{Dependence on the tolerance }
\label{sec:tolerance-time}
\indent

We have examined the values $\theta$ = 0.5, 0.6, 0.7, 0.8, 0.9, 1.0
and 1.1. As expected, the time per step decreases as the tolerance
increases, and that for all values of $n_{crit}$ and of $N$. This
decrease, however, is relatively less important than that found for a
standard tree code (Hernquist 1987). Furthermore, it is
not equally important for all values of the
tolerance, and in particular is relatively small for $\theta > 0.8$.
This is clear from Fig.~\ref{time_n},
where the curves corresponding to $\theta \geq 0.8$
cluster together.

\subsection{Dependence on the clustering of the points }
\label{sec:cluster-time}
\indent

All the tests given above, as well as most given so far in the
literature, concern a Plummer sphere distribution. Yet conventional
wisdom says that the performance
of the tree code should depend heavily on how the points are
distributed. In order to test this we measured the CPU per time step
in configurations of variable clumpiness.

For this we considered 10 points (which can be considered as galaxy 
centers) randomly distributed in a given
volume $D^3$ and around each a
number of points distributed with a radial density profile
$exp(-\alpha r)$ (which can be considered as representing a 
``galaxy''). For large values of $\alpha D$ the
configurations will of course be more concentrated around the 10
centers, than for small values of $\alpha D$. We find that the CPU time
necessary for one time step decreases with increasing clustering, but
that the effect is relatively small. Thus for $N=$ 25~000, $\theta$ = 0.5, 
$n_{crit}=$ 8~000, $D=$20 and $\alpha$~=
0.5, 1., 2. and 5. we have respectively 0.97, 0.90, 0.82 and 0.79 
seconds. For $N=$ 500~000 and
the same values for the other parameters we have 31.23, 31.09, 30.76
and 29.78 seconds.
This trend goes against the conventional wisdom that more clustered
configurations should necessitate longer CPU times, but the
differences can be easily understood in terms of the differences
between the standard Barnes \& Hut tree code and our GRAPE version of
it. Indeed in the Barnes \& Hut tree code a more centrally concentrated
configuration means that the tree descent has to go deeper and is
therefore more time-consuming. This, however, is not the case for our
version of the tree code since for the particles in the same block 
we use direct
summation and therefore it makes no difference whether they are
centrally concentrated or not. On the other hand for a system with a large
value of $\alpha D$ ``galaxies'' will be very centrally concentrated and
therefore will need few subdivisions for the tree when seen from a
point in another ``galaxy''. Thus more clustered configurations should 
necessitate somewhat less CPU time than less clustered ones, as was
indeed seen in the numerical examples.

\section{Timing of GRAPE-3AF}
\label{sec:timing}
\indent

\subsection{The effect of the front end } 
\label{sec:front end}
\indent

In order to see how much of a difference a faster front end can make to
the speed of
 our simulations we have repeated the beginning of two simulations
with the front ends at our disposal and compared the running time. The first
simulation is a
direct summation case with 65 400 particles. It is run ki64 in the 
notation of Athanassoula, Makino \& Bosma (1997) and represents a 
compact group of five identical Plummer galaxies distributed 
according to a King ${\Psi}=1$ law. 87.5\%
of the mass of 
the system is in a common halo which follows a similar mass distribution and 
encompasses the whole group. We 
find that 
the Sparc 10/512, the Ultra 1 and the Ultra 2 take 6.26, 5.90 and
5.89 
seconds 
per time step respectively. The differences are small, due to the fact
that most
 of the work in the direct summation case is done by the GRAPE system.

Our second example is the evolution of a barred galaxy with a live
halo, represented by 120~000 particles in total. In this case we
follow the 
evolution with a tree code with an opening angle of 0.6 and no
quadrupole 
terms. We find that the Sparc 10/512, the Ultra 1 and the Ultra 2
take 17., 
8.2 and 7.9 seconds per time step respectively. In this case the difference 
between the performances of the three machines is more important than 
in the previous
example. A more meaningful comparison, however, would involve only the
time spent on the front end, so that we have to subtract from the
total time that used for the force calculations on GRAPE and the 
communication time between the host and the boards. About 12 Mbytes of
data have to be transferred between the host and GRAPE at every time
step. Since the effective transfer rate of the transfer box is about 3-4
Mb/sec, the data transfer would take about 3-4 seconds. Roughly extrapolating
from Makino (1991) we find the length of the interaction list to be around
3~000, which, for the 120~000 particles in our system, gives 3.6
$10^8$ interactions. Taking into account that our 5 board GRAPE-3AF system has 
a speed of 8~$10^8$ interactions/second, we find that the time spent for
the calculations is less than a second. Thus the total time spent for
communications and calculations on GRAPE is of the order of 4 seconds
and most of it can be accounted for by the communications.
This gives us that the times spent on the front end are 3.9, 4.2 and 
12.9 seconds for our
Ultra 2, Ultra 1 and the Sparc 10/512 respectively. The ratio of the 
times on the
Ultras is in good agreement with the ratio of their clock speeds. For
a comparison with the Sparc 10/512 we have to use specfp values, so
that the comparison is not as straightforward. Nevertheless the
ratios agree to better than a factor of two.

\begin{figure}
\includegraphics{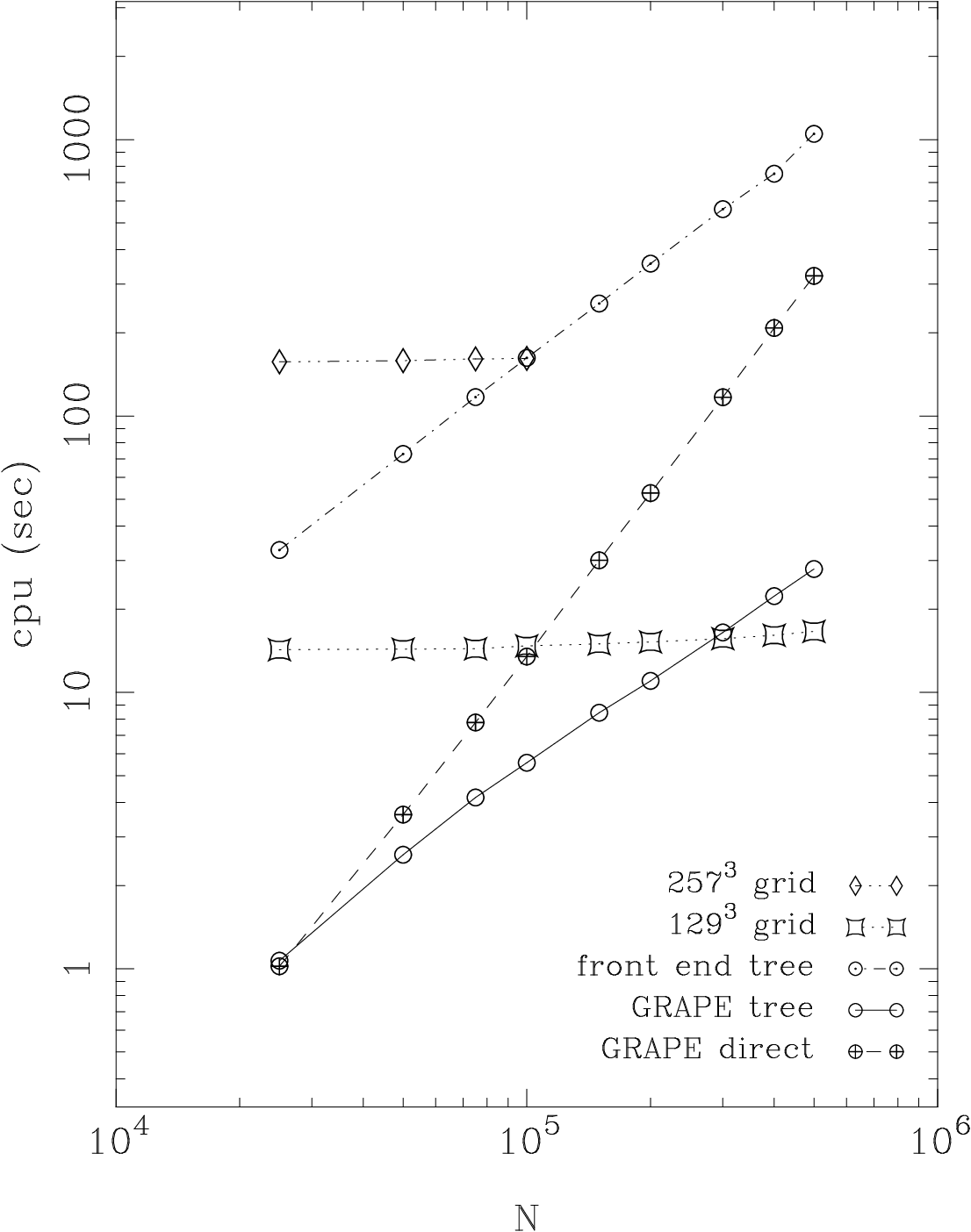}
\vspace{11.5cm}
\caption{
CPU time necessary for one time step, as a function 
of the number of
particles in the system. Results are shown for direct summation on
our GRAPE-3AF system (dashed lines and circles with crosses), 
the tree code on our GRAPE-3AF system with
a tolerance of 0.7 (solid line and open circles), a tree code on
the front end and with the same value of the tolerance (dot-dashed lines 
and circles with a dot) and a three
dimensional cartesian grid code with a 129x129x129 grid (dotted lines
with lozenges) and with a 257x257x257 grid (dashed-dotted-dotted lines
and diamonds).
}
\label{alltimes}
\end{figure}

\subsection{GRAPE-3AF compared to other potential solvers }
\label{sec:timing-general}
\indent

Fig.~\ref{alltimes} compares the CPU times necessary for one time step
as a function of the number of particles in the system and different
codes. The GRAPE timings were obtained using our GRAPE-3AF system with
direct summation and tree code respectively. For the tree code on the
front end we used the version available in NEMO, which is due to
Josh Barnes. For the cartesian grid code we used Jerry Selwood's code,
which uses Richard James's potential solver. 

The break-even point between direct summation and tree code on our
GRAPE-3AF system is around 25~000 particles. For a higher number of
particles the tree code is faster. This number of course depends on the
ratio of the CPU performance of GRAPE and of the front end and would
thus be different for another front end or a different number of
boards. 

Comparison of the NEMO scalar tree code and the GRAPE tree code shows that
the GRAPE tree code is about 30 times faster for a small number of
particles and about 40 times faster for larger numbers. We have not
searched for the tree code which would run fastest on our front
end. The version we are using, however, has the advantage of not being
optimised for a vector machine, which would unnecessarily hamper its
performance on a scalar machine. Also the version of the tree code which is
running on our GRAPE-3AF system is a rather straightforward implementation of
a previously existing version and it should be possible to achieve 
considerable gains in performance by rewriting the code according to
the needs of the GRAPE boards. Such a work is in progress.

For the cartesian grid code we have considered two different
resolutions, a high resolution grid of 257x257x257 and a low
resolution one of 129x129x129. We have not been able to extend the
tests for the high resolution grid code beyond 100~000 particles,
because of memory limitations, since grid codes require considerable
memory allocations. Nevertheless we can deduce that the high resolution
grid code runs considerably slower than our GRAPE tree code for all
number of particles tested and even slower than our 
direct summation GRAPE for less than of the order of 40~000
particles. On the other hand the low resolution grid code runs faster
than the direct summation for more than 100~000 particles and faster
than the tree code for more than 300~000 particles. 

It is thus clear from the above diagram that our GRAPE codes, both
direct summation and tree code, give very high performances, and can be
used with a very large number of particles, the last statement being
particularly true for the tree code.

\section{Accuracy of direct summation on GRAPE-3AF}
\label{sec:n2-accu}
\indent

\begin{figure}
\includegraphics{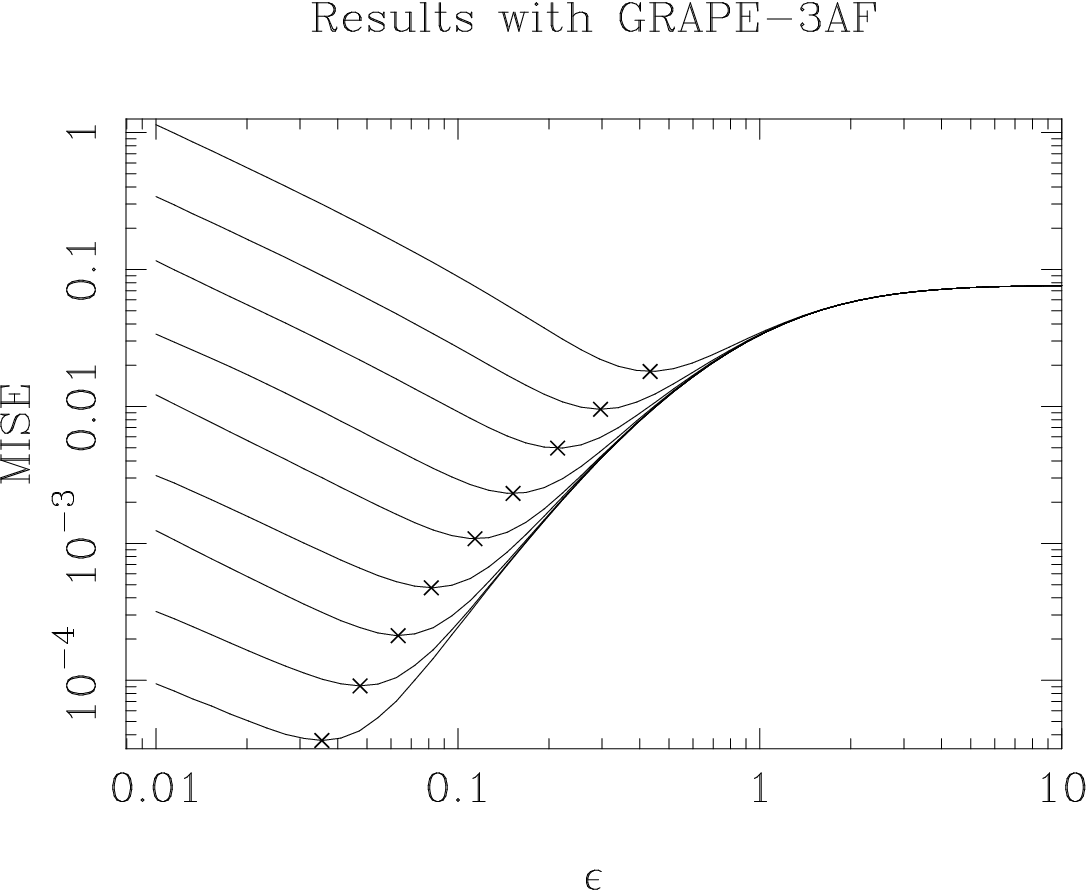}
\vspace{8.cm}
\caption{
$MISE$ as a function of the softening $\epsilon$ for a Plummer
sphere. From top to bottom the 
curves correspond to 
$N = 30, 100, 300, 1~000, 3~000, 10~000, 30~000, 100~000, 300~000$, 
where $N$ is the
number of particles in the realisation of a Plummer sphere.
The position of a minimum error along a line for a given $N$ is marked
by
an X, and the corresponding $\epsilon$ value is the optimum softening 
$\epsilon_{opt}$
for this number of particles.
}
\label{mise}
\end{figure}

\subsection{Accuracy of the force calculation }
\label{sec:n2-mise}
\indent

\begin{figure}
\includegraphics{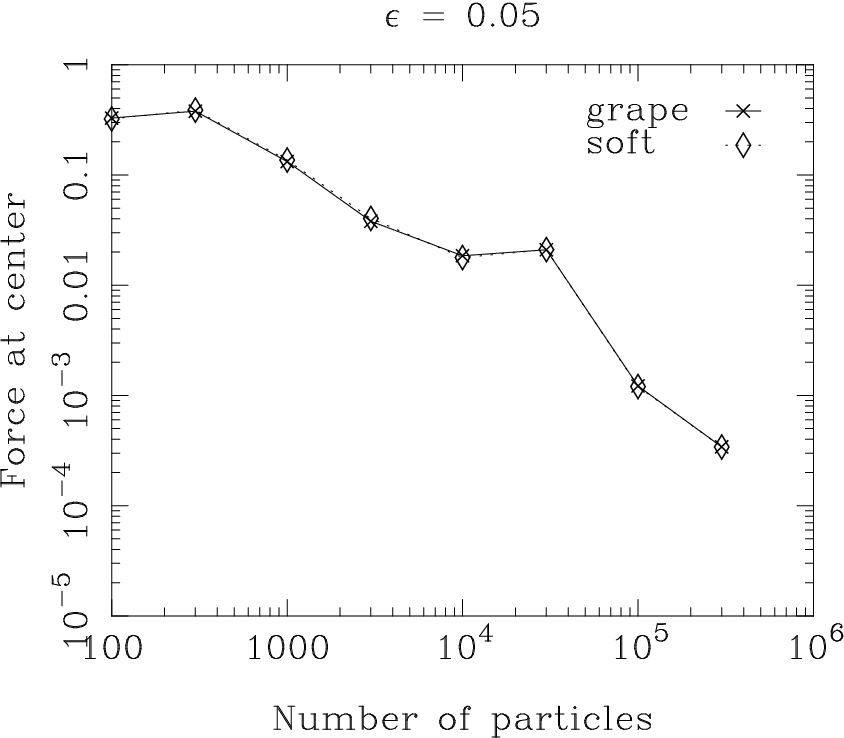}
\vspace{7.5cm}
\caption{
Amplitude of the force at the center of a Plummer sphere as a function 
of the number of points $N$ in the realisation for direct summation on 
a work station with 32 bits accuracy (solid lines and diamonds) and
on GRAPE-3AF (dashed lines and crosses).
}
\label{plum_force}
\end{figure}

Merritt (1996) and Athanassoula \tal (1997, in preparation) discussed the
value of the softening 
($\epsilon$) which gives the best approximation of the force due to a given 
density distribution. For this they use the quantity

$$MISE = <\int \rho ({\bf x}) |{\bf F}-{\bf F}_{true}({\bf x})|^2 d{\bf x}>$$

\noindent
and a density distribution corresponding to a Plummer sphere of unit
mass and unit scale length. We repeated the exercise using 
direct summation on GRAPE-3AF. Since a Plummer distribution is
spherically symmetric the above simplifies to a one-dimensional
integration along a radius. For this we use the alternative extended
Simpson's rule (Press \tal 1988), which has an accuracy of 
${\cal O}(N^{-4})$, and 100 points along the line of integration. 
The upper limit of the integration was taken to be $L=20b$, where 
$b$~=~1 is the 
scale length of the Plummer sphere. This radius contains more than 99\% 
of the mass of the Plummer model, while there the density has fallen 
at that point to 
roughly $3 \times 10^{-7}$ of its central value. The number of
realisations was taken to be $6\times 10^6 /N$. More details on
this calculation are given in
 Athanassoula \tal (1997, in preparation). The results 
are shown in 
Fig.~\ref{mise}, and coincide within the mean errors with the 64 
bits direct summation results. This could, at first sight, sound at odds with 
the fact that GRAPE-3AF is a low precision machine. Nevertheless one should 
keep in mind that the error in GRAPE-3AF comes from round-off and 
thus can be considered random. To illustrate this we calculated 
the force at the center
of a Plummer sphere represented by $N$ points. In the continuum limit
this force should be equal to zero, so its amplitude is a measure of
the error in the calculations. We plot it in Fig.~\ref{plum_force} as
a function of the number of particles $N$ in the realisation, for
direct summation with GRAPE or using 32 bit precision on the front
end. It is interesting to note that the results of the two methods are the 
same. This, as already mentioned above, is due to the fact that the
errors for GRAPE are random and thus cancel out when we sum the
force contributions from the particles in the realisation. This argument,
together with the sharp decrease in the error as the number of
particles is increased, shows clearly that, in order to increase the
accuracy of a simulation, one should increase the number of particles
rather than the accuracy of the force calculation, provided, of course
that the errors in the force calculations are not systematic. 
In this we agree with the results found, in a
different way, by Hernquist, Hut \& Makino (1993) and Makino (1994). 

\begin{figure}
\includegraphics{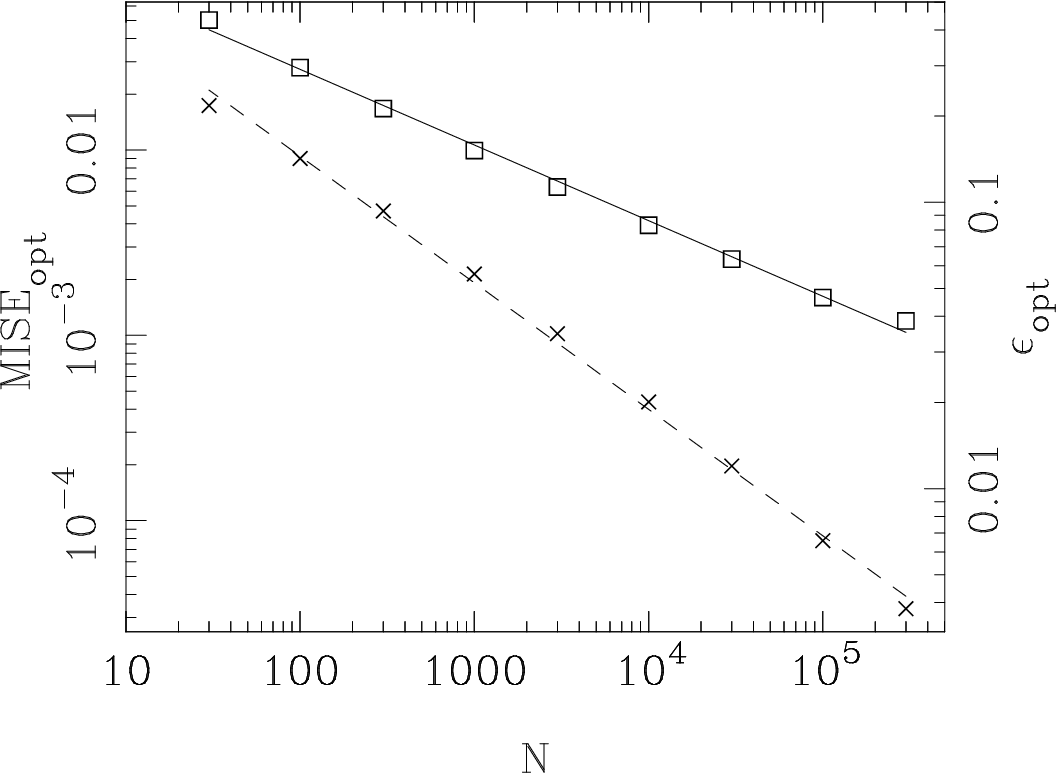}
\vspace{7.0cm}
\caption{
Optimum softening length, $\epsilon_{opt}$ as a function of number of 
particles (squares and solid line; scale on the ordinate on the right)
and corresponding $MISE$ values (crosses and dashed line; scale on the
ordinate on the left).}
\label{mise_accu}
\end{figure}

We calculated from curves such as shown in  Fig.~\ref{mise} 
the minimum $MISE$ value for a given number of particles $N$
simply by fitting a second order polynomial to the three points with
the lowest $MISE$ values. We thus found the 
optimum softening length,
$\epsilon_{opt}$, and the corresponding $MISE$ value as a function of
the number of particles $N$ and we display them in
Fig.~\ref{mise_accu}.  
Both can be well represented
by power laws:

\begin{equation}
$$\epsilon_{opt}=0.98N^{-0.26}$$
\end{equation}

\noindent
and

\begin{equation}
$$MISE=0.22N^{-0.68}$$
\label{eq:mise_N}
\end{equation}

There is, however, an indication the $log(MISE)$ is not a linear
function of $logN$, but that a second order polynomial would be a better
fit. Thus for small numbers of particles the error increases with
decreasing
$N$ less fast than for large number of particles. For that reason the
exponent of eq. (\ref{eq:mise_N}) will depend somewhat on the range of 
$N$ considered.

Our results agree nicely with those of Merritt (1996). In particular
we find the same exponent for the dependence of $\epsilon_{opt}$ on $N$,
while we find that the
dependence of $MISE$ on $N$ is somewhat steeper, which, taking into account
the effect discussed in the above paragraph,
can be understood, since our results extend to higher numbers of
particles. This point will be taken up again in section \ref{sec:tree-mase}.

\subsection{Simulations of a Plummer sphere }
\label{sec:n2-sim}
\indent

\begin{figure}
\includegraphics{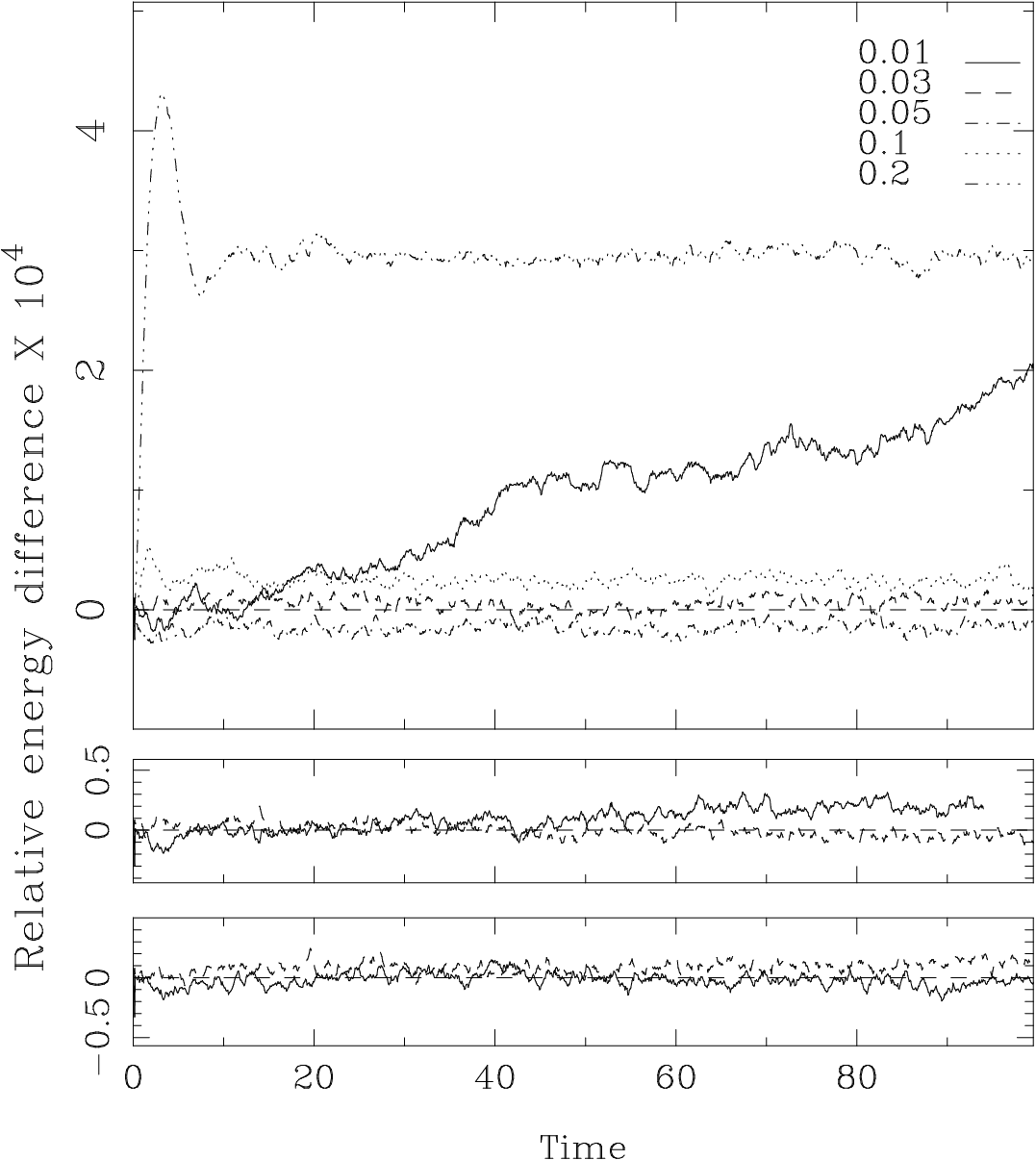}
\vspace{9.5cm}
\caption{Relative energy difference, $(E(t)-E(0))/E(0)$, as a
function of time for five simulations of the evolution of a Plummer 
sphere realisation with 100~000 particles. Each evolution corresponds to
a different softening, 0.01 (solid line), 0.03 (dashed line), 0.05 
(dash-dotted line), 0.1 (dotted line),
and 0.2 (dash-dot-dot-dot line). A horizontal dashed line at zero
relative energy difference has been plotted to guide the eye.
The upper panel corresponds to a time step of 1/64, the middle one to
1/128, and the lower one to 1/256.}
\label{100kplum_energy}
\end{figure}

In the above we discussed the effect of the softening on the
calculated value of the force. However the accurate representation of
the force is only one of the
aspects to be taken into account in numerical 
simulations. Furthermore the largest contribution to the $MISE$ or
$MASE$ comes from the immediate neighbourhood of each particle or
point, while classical theory of two-body relaxation tells us that it
is the contribution of distant particles that dominates the relaxation
effect. For this reason 
we evolved a 100 000 particles Plummer sphere
using direct summation on the GRAPE-3AF boards and five different 
values of the
softening, namely 0.01, 0.03, 0.05, 0.1 and 0.2 and checked 
energy conservation. Our Plummer sphere is
in virial
units, i.e. $G=1$, $M=1$ and $4E=-1$, where $G$ the gravitational
constant, $M$ the total mass of the Plummer sphere and $E$ its total
energy (kinetic plus potential). This gives a scale length of
$b=3 \pi / 16$. The softening value for which the
force is optimally described for 100 000 particles is 
$0.047*3\pi/16=0.028$, i.e. within the range of values tried and very
near the second value 0.03.
The upper panel of Fig.~\ref{100kplum_energy} shows the relative
energy difference as a function
of time for the five simulations evolved with a time step of 
$\delta t$~=~1/64~=~0.015625. In order to lessen the noise
and bring out trends more clearly we applied nine point sliding means to the
data. We see that the case with the highest softening, where the
force calculation contains a considerable bias, starts out of
equilibrium and within the first few time steps readjusts. Thus the
energy evolves fast to a value much different from the initial one. 
The three intermediate
cases have good energy conservation, the best one being the case with
a softening of 0.03, which is the value nearest to the one predicted
by the minimum of the $MISE$ as a function of $\epsilon$ curve
(cf. Fig.~\ref{mise}).  
For the case with the smallest softening the energy is badly
conserved, showing a steady increase with time. This, however, is not
due to the value of the softening, but to the combination of the value
of the softening and that of the time step. Indeed we used
a ratio of $\epsilon/\delta t$~=~0.64 while for our Plummer sphere
$<u^2>^{1/2}$~=0.7. We thus repeated the simulations for half and a
quarter of the time step, i.e. $\delta t$~=~1/128 and 1/256. The
results are plotted in the lower panels of Fig.~\ref{100kplum_energy},
and show that for a sufficiently small time step the energy can be
well conserved. 

Since both a good conservation of the energy and an accurate
representation of the force are important for high quality numerical
simulations the choice of the appropriate $\epsilon$ is a complex
matter. The softening that gives the most accurate representation 
of the force can be found by calculations such as those in the
previous subsection. On the other hand a good energy conservation
depends on both the softening and the time step. However a small value of
the softening imposes the choice of a small time step, which in turn
imposes higher CPU time requirements.

\section{Accuracy of the tree code on GRAPE-3AF  }
\label{sec:tree-accu}
\indent

\subsection{Accuracy of the force calculation }
\label{sec:tree-mase}
\indent

In order to quantify the accuracy of the tree code and compare it to
that of direct summation we have used the quantity $MASE$, introduced
by Merritt (1996, cf. also Athanassoula \tal 1997).

$$MASE = <\frac{1} {N} \sum_{i=1}^{N}|{\bf F}_i-{\bf F}_{true}({\bf x}_i)|^2>$$

This is similar to $MISE$, used in section \ref{sec:n2-mise}, but the
force is now calculated on all particles in the configuration, rather
than at some points on a line. Thus it is much more
time-consuming to calculate than $MISE$, but this does not pose in our
case any problem, since the tree code on GRAPE is very fast. For the
same reason, and in order for the range of values of $N$ to be
comparable to what we can use for simulations with the corresponding 
code on GRAPE,
we have examined a different range of values of $N$ than what was used
for direct summation. Namely we
have not considered values less than 30~000, which would,
anyway, be meaningless with the adopted values of $n_{crit}$, while we extended
the upper end for $N$ to 1~000~000, a value which as we saw in section
\ref{sec:timing-general}, can be used without problem on our GRAPE 
configuration.

\begin{figure}
\includegraphics{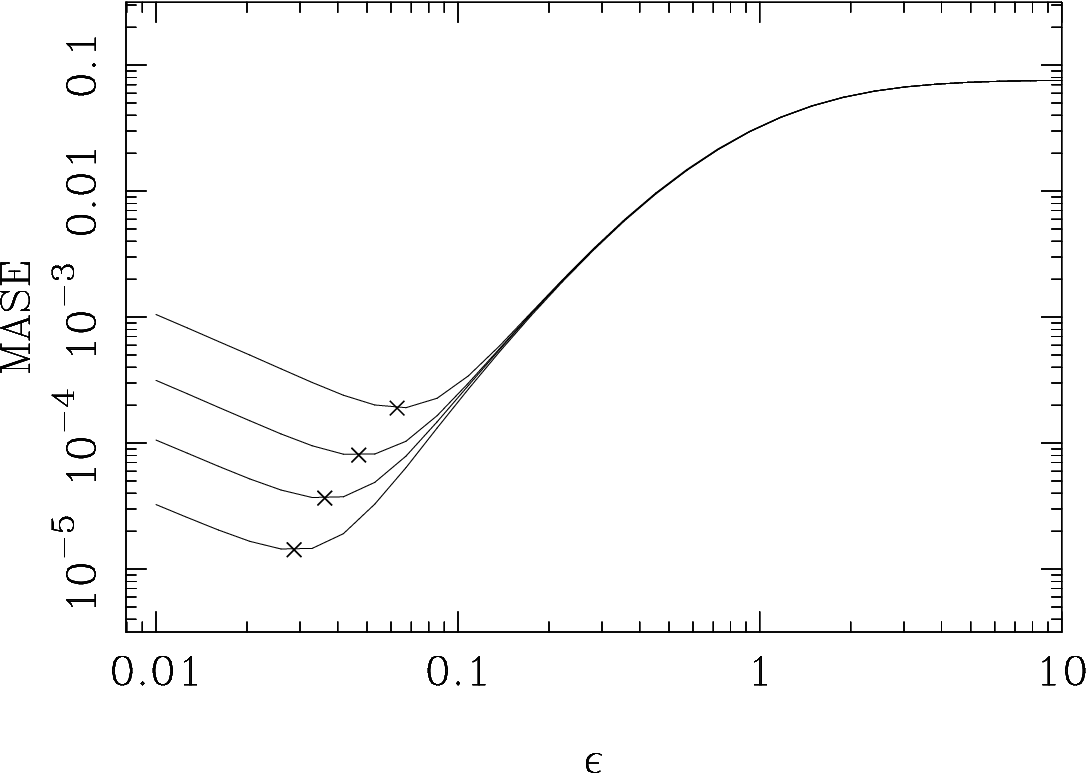}
\vspace{6.0cm}
\caption{
$MASE$ as a function of the softening $\epsilon$ for a Plummer
sphere and a tree code with $\theta$ = 0.5 and $n_{crit}$~=~4~000. 
From top to bottom the 
curves correspond to 
$N =$ 30~000, 100~000, 300~000, and 1~000~000,
where $N$ is the
number of particles in the realisation of the Plummer sphere.
The position of a minimum error along a line for a given $N$ is marked
by
an X, and the corresponding $\epsilon$ value is the optimum softening 
$\epsilon_{opt}$
for this number of particles.
}
\label{tree_mase}
\end{figure}

\begin{figure}
\includegraphics{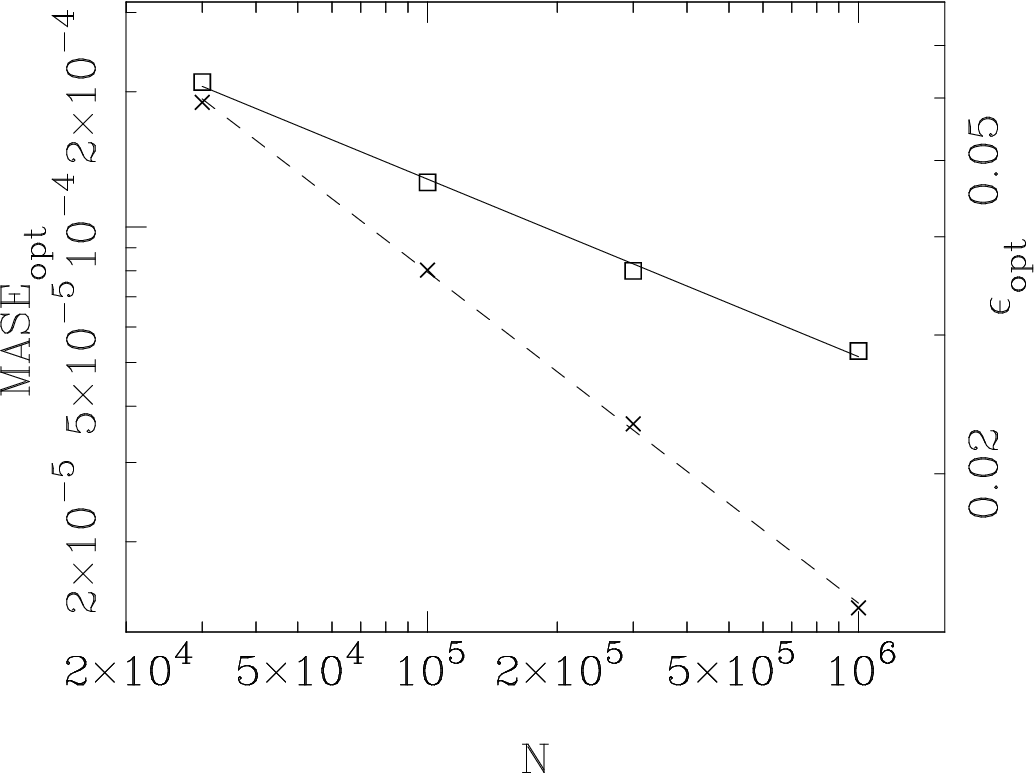}
\vspace{6.5cm}
\caption{
Optimum softening length, $\epsilon_{opt}$ as a fuction of number of 
particles (squares and solid line; scale on the ordinate on the right)
and corresponding $MASE$ values (crosses and dashed line; scale on the
ordinate on the left).}
\label{mase_accu}
\end{figure}

The results obtained with $n_{crit}$~=~4~000 and $\theta$ = 0.5 are given in
Figures \ref{tree_mase} and \ref{mase_accu}. 
The first thing to note by comparing
Figures \ref{mise} and \ref{tree_mase} is that the accuracy of the
tree code is comparable to that of direct summation. This can be
understood since the main source of error in $MISE$ or $MASE$ comes
from the nearby particles and, since these are treated by direct
summation in both cases, we get comparable values for the two methods. 

As shown in Fig. \ref{mase_accu} both the optimum softening length
$\epsilon_{opt}$ and the corresponding $MASE$ value, $MASE_{opt}$, can
be represented as power laws of the number of particles $N$

\begin{equation}
$$\epsilon_{opt}=0.63N^{-0.22}$$
\end{equation}

\noindent
and

\begin{equation}
$$MASE_{opt}=0.38N^{-0.73}$$
\label{eq:mase_N}
\end{equation}

These numbers are considerably different from those found in section  
\ref{sec:n2-mise}. The difference, however, is not due to a difference
between the two codes but to a difference between the range of number
of particles considered in the two cases. This is made clear in table
1, where we give the values of the exponent for
$\epsilon_{opt}$ (column 3) and $MASE_{opt}$ (column 4), together with
the corresponding range of particle numbers $N$ (column 2). The first
line repeats the values from Merritt (1996) and the second one those
of section \ref{sec:n2-mise}. For the third and fourth line we have
considered separately two ranges of particle numbers, both for direct
summation. The fifth line gives our values for the tree code. Finally
the sixth line gives the asymptotic values obtained for $N\rightarrow
\infty$, as calculated in the Appendix. We note that indeed as the
range of particle numbers becomes higher the exponents approach their
asymptotic limit.

\begin{table}
\label{tab:index}
\caption{Exponents of power laws}
\begin{center}
\begin{tabular}{lccc}
\hline
Case & Number of & exp. & exp. \\
 & particles &  $\epsilon_{opt}$ &
$MASE_{opt}$  \\
\hline
Merritt (1966) & 30 - 3 $\times 10^4$ & -0.28 & -0.66   \\
Direct summation & 30 - 3 $\times 10^5$ & -0.26 & -0.68  \\
Direct summation & 30 - 3000& -0.29 & -0.62  \\
Direct summation & 1 $\times 10^4$ - 3 $\times 10^5$ & -0.23 & -0.76  \\
Tree code & 3 $\times 10^4$ - 1 $\times 10^6$ & -0.22 & -0.73   \\
Asymptotic & $\infty$ & -0.2 & -0.8   \\
\hline
\end{tabular}
\end{center}
\end{table}

We also calculated $MASE$ values for different values of the tolerance
and of $n_{crit}$. The results came out as expected, i.e. the force
calculations were more precise for smaller tolerances or larger
$n_{crit}$, but the effects were small. The fact that, contrary to the
standard tree code, the effect of the tolerance on the accuracy is
small, can be easily explained by the fact that the largest
contribution to $MISE$ or $MASE$ comes from relatively nearby
particles, for which the force in our tree code is anyway calculated by
direct summation. This argues that, in order to get a higher accuracy,
it is preferable to increase the number of particles rather than to 
decrease the tolerance.

\subsection{Simulations of a Plummer sphere }
\label{sec:tree-sim}
\indent

We repeated the five simulations of the previous section, using this
time the tree code instead of direct summation. The results were quite
satisfactory, although the tree code simulations conserved energy
somewhat less well than the direct summation ones, as could be
expected. The run of the energy with time can be mentally decomposed
into some global trend on which is added some noise. The average
relative value of this noise was of the order of 4 parts in $10^5$ and
did not seem to depend on the value of the softening. The global trend
gave an energy conservation of 4 parts in $10^5$ for $\epsilon = 0.01$ and
a time span $\Delta t=100$, and less than 2 parts in $10^5$ for the remaining
values of the softening.

\section{Long-term evolution of a barred galaxy }
\label{sec:bar}
\indent

\begin{figure*}
\includegraphics{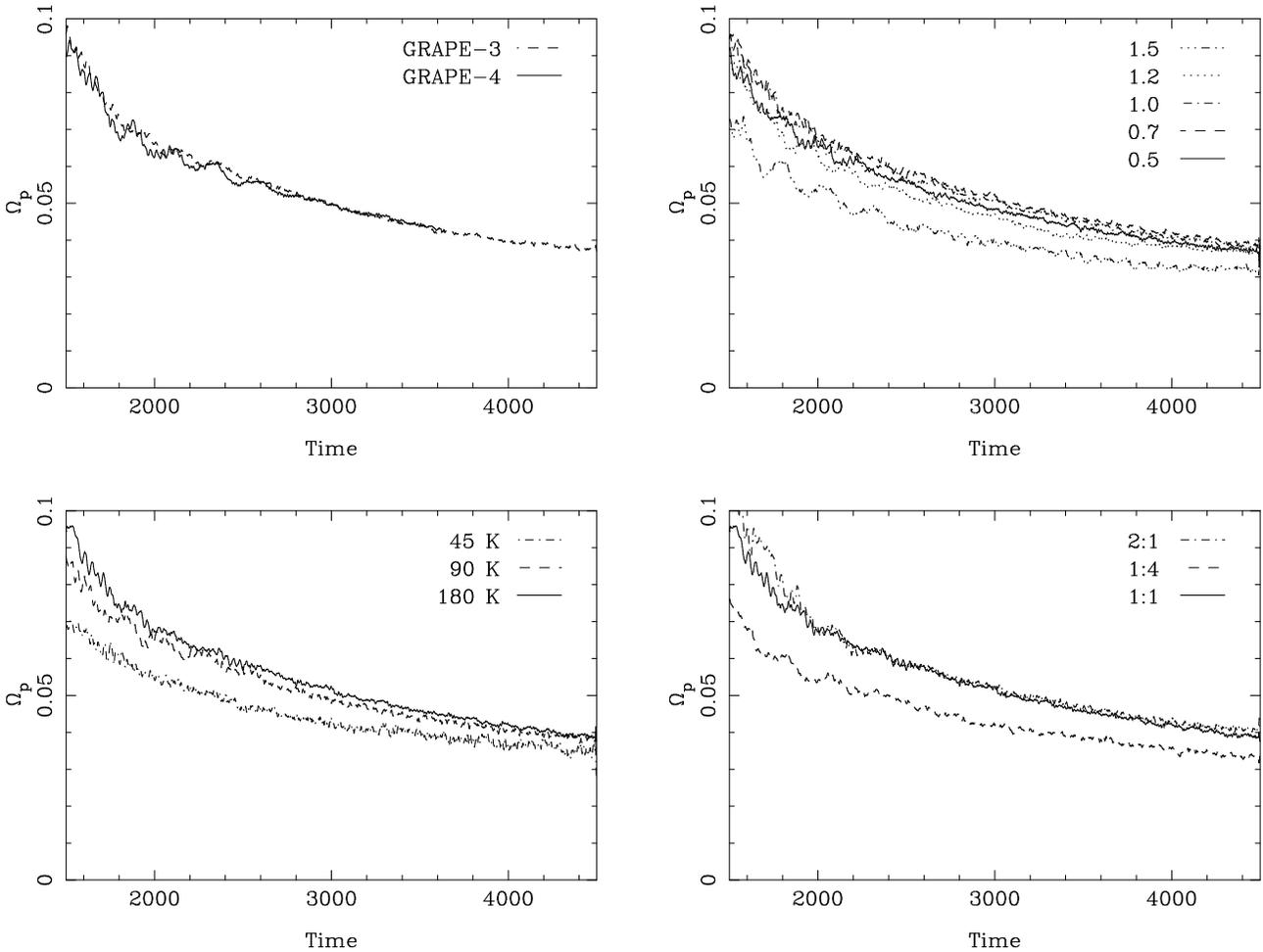}
\vspace{14.0cm}
\caption{
Pattern speed of the bar as a function of time. The upper left panel
compares results obtained with GRAPE-4 (solid line) and GRAPE-3AF
(dashed line). The upper right panel compares results
obtained with the tree code and tolerances of 0.5 (solid line), 0.7
(dashed line), 1.0 (dash-dotted line), 1.2 (dotted line) and 1.5 
(dash-dot-dot-dotted line). The
lower left panel compares simulations with the tree code and a 
tolerance of 0.7 for 180 000 particles (solid line), 90 000 particles
(dashed line) and 45 000 particles (dash-dotted line). The lower right panel
again refers to simulations with a tree code and a tolerance of 0.7. The
solid line corresponds to a simulation with equal mass particles in
the disc and in the halo, the dash-dotted line to a simulation where the disc
particles are twice as massive as those of the halo and the dashed line
to a simulation where the mass of the particles in the halo is four times as
big as that of the particles in the disc. 
}
\label{omegap}
\end{figure*}

In this section we discuss the evolution of the same initial conditions with
different methods, in order to be able to assess the effect of the
code on the results. For this we use initial conditions corresponding
to a bar-unstable disc galaxy, where
the halo is described by 120~000 particles and the disc by 60~000, and
evolve it using direct summation on GRAPE-4 (for a description of this
high accuracy machine, see Makino et al. 1997), direct summation on
GRAPE-3AF, 
and tree code on GRAPE-3AF with 5 different values of the tolerance.
Quantities that allow us to make quantitative comparisons between the
different results are the pattern speed and the amplitude of the bar.
For this we simply measure the phase and amplitude of the $m=2$  
component as a function of radius on-line in all three cases 
with the help of the same software. This allows us to calculate the
pattern speed from the smoothed time derivative of the phase, averaged
over the radii where the bar is best defined. 
The results are displayed in Fig.~\ref{omegap}. The
observed decrease is due to the exchange of angular momentum between 
the bar on the one hand and the disc and halo on the other, and its
interpretation, as well as the discussion of its importance, have been 
the subject of many papers (e.g. Weinberg 1985, Hernquist \& Weinberg
1992, Little \& Carlberg 1991, Athanassoula 1996, Sellwood \&
Debattista 1996). Here we
will only be interested in how well the results of the various codes
agree with each other. In the upper left panel we compare the
results of GRAPE-3 with those of GRAPE-4, and find excellent agreement.
 A comparison between the GRAPE-3 results
and those with a tree code and a tolerance of 0.5 (not shown here) are 
also very satisfactory. In the upper right panel we compare 
tree code results with tolerances of 0.5, 0.7, 1.0, 1.2 and 1.5. 
The values obtained with the two biggest values of the tolerance
are considerably smaller than those obtained with the other
values. The lower left panel compares the results obtained in all
cases with a tree code with a tolerance of 0.7 and different number of
particles, 180~000, 90~000 and 45~000 respectively. We note that the
values of the pattern speed obtained for 45~000 particles are
considerably lower than the others, while the
other two are relatively close, arguing that they are converging and
therefore that 180 000 particles are
sufficient for such a simulation.

\begin{figure*}
\includegraphics{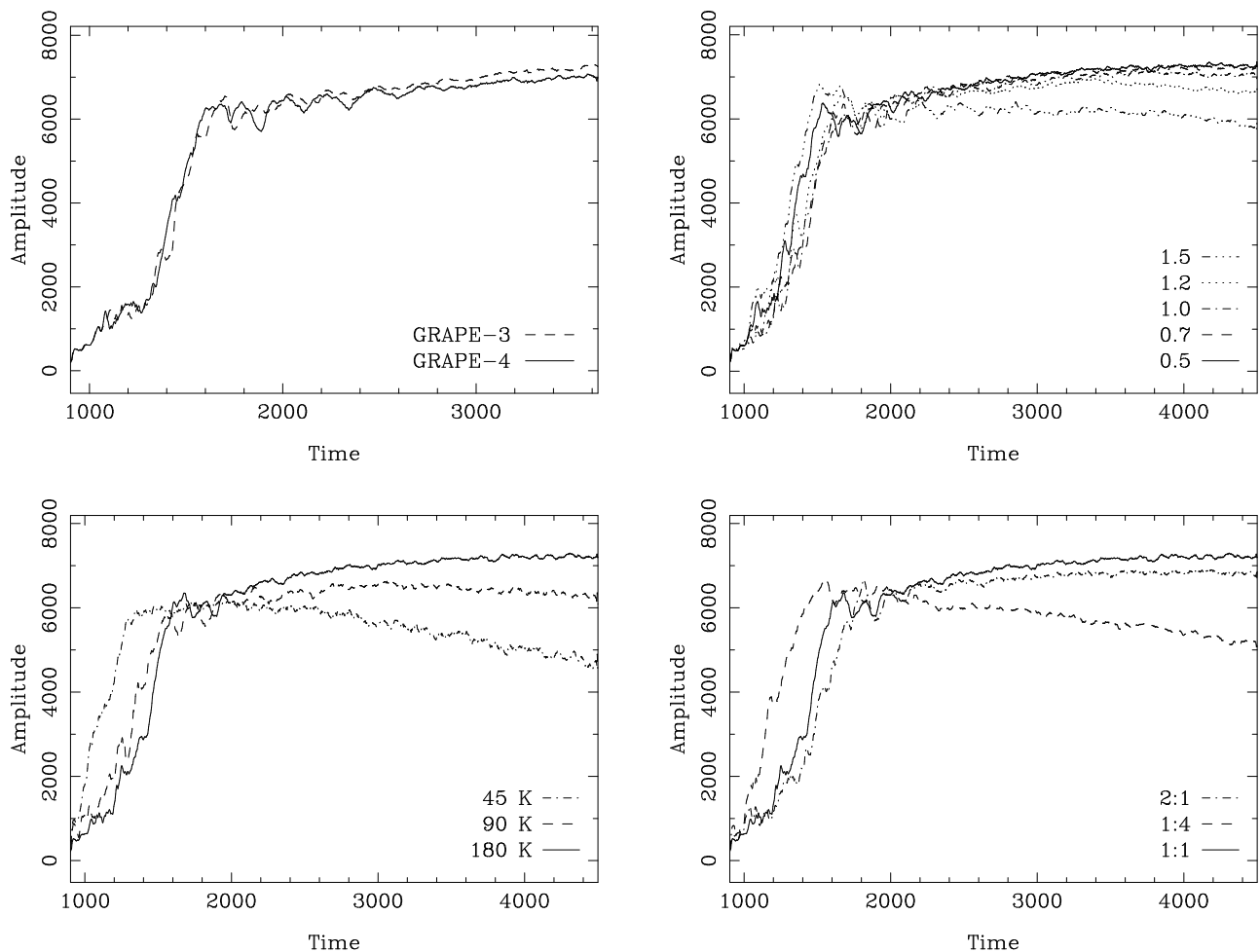}
\vspace{14.0cm}
\caption{
Measure of the $m=2$ component of the density as a function of
time. The layout and the various line styles are the same as for the
previous figure. 
}
\label{baramp}
\end{figure*}

Finally the lower right panel
compares the results in the case when all the particles in the
simulation are not of the same mass. In one case we have considered
particles in the disc which are twice as massive as those in the halo
and in the other particles in the halo which are four times as
massive as those in the disc. This last case gives results which are
considerably lower than those of the other two. The ``trick'' of using
more massive and therefore fewer particles for the halo, in order to 
have more particles in the
disc, has been often used in numerical simulations, but the above
results show that such simulations should be interpreted with
caution. Of course one can argue that, since we do not know what the
halos are made of, we do not know what value to use for the ratio of
the mass of the halo particles to that of the disc
particles. Nevertheless it should be kept in mind that the adopted
value could influence considerably the results. For example in merging
simulations heavier halo particles could sink faster to the center
than lighter ones, and thus lead to a more concentrated final halo.
A measure of the bar amplitude is given in Fig.~\ref{baramp} as a
function of time. The
layout of the different panels is the same as that of
Fig.~\ref{omegap}. We again note that the GRAPE-3 and GRAPE-4 results
agree very well. Also the bar amplitude is smaller in the cases with
higher tolerance values, as could be expected since higher tolerance
gives more smoothing. The same thing is seen also for simulations with
a lower number of particles. Finally, as for the previous figure, the
mass of the particles in the halo influences considerably the
results. 

\section{Summary }
\label{sec:summary }
\indent

In this paper we present the Marseille GRAPE-3 systems and discuss
their possibilities and limitations. At present we have a system with
5 GRAPE-3AF boards linked to an Ultra 2/200, and a single GRAPE-3A board
linked to a Sparc 10/412 workstation. Since the N-body problem allows
for a straightforward parallelisation it is easy to take full advantage
of a five board system. Further time can be gained if the data
transmission to some chips overlaps in time with calculations on other
chips.

In order to keep the storage requirements to a reasonable level it is
mandatory to use on-line analysis. The complication here, compared 
to simulations run on workstations, is that GRAPE is very
much faster than the front end, so it is unreasonable to have it wait 
while the slow front end does the analysis. The problem is solved if
the workstation has a second processor which is assigned to do the
on-line analysis. In cases where this analysis involves heavy
calculations that can be done on a GRAPE, we have found it useful to
use our slower GRAPE-3A system for that task. The on-line analysis we
perform includes the production of short movies, giving visual
information on the simulation, as well as the calculation of several
quantities for the different components in isolated galaxies or the
different galaxies in pairs, groups or clusters. We give increasingly
more importance to the on-line analysis and its proper planning can
take as much time as the preparation of the initial conditions.

On our GRAPE systems we run two kinds of 
simulation software, direct summation
and the tree code, and we give here a description of each. 
In the case of the tree code particles are divided in blocks
with a common interaction list and then direct summation is used over
this list.
We analyse the performance of the tree code and
how this depends on various parameters such as the number of particles in
a block, the total number of particles, the tolerance and
the clustering of the points. We thus find the optimum number of
particles per block for our system to be of the order of 7~000 to
8~000. Contrary to conventional wisdom, we find that less clustered
configurations take longer CPU times and this can be explained by the
differences between the conventional Barnes-Hut tree code and our
version which uses direct summation between particles in the same
block. 

A faster front end brings little improvement to the performance of the
direct summation code, since most of the time is taken by calculations
on the boards. The opposite is true for the tree code, where a faster
front end can make considerable difference, as we show by 
analysing the relative
times on the front end and on the boards. 

We then compare the performances of our GRAPE system with that of a
tree code and a cartesian grid code run on the front end. As expected,
we find that the GRAPE tree code is a very significant improvement
compared to the front end version, but even direct summation goes much
faster than the front end tree code, at least for a number of particles
up to a million or more. Both our codes are also considerably faster
than a high resolution (257x257x257) grid code. Only the low
resolution (129x129x129) cartesian grid code beats our tree code for
more than 300~000 particles. Furthermore, a considerable improvement in the
performance of our tree code can be expected if the code is rewritten
according to the specifications of the GRAPE boards.

A powerful potential solver can be useless if it does not have 
sufficient accuracy. For this reason we analyse extensively the
accuracy of both direct summation and tree code. For the accuracy of
the force calculation we use the concepts of $MISE$ and $MASE$
introduced by Merritt (1996), which allow a clear estimation of how
accurately the force is calculated. We find that the forces are as
accurate as when full accuracy is used on a front end. The reason is
that errors in the force calculations on GRAPE are due to round-off
and are thus purely random. Thus they cancel out when we sum the force
contributions from a large number of particles. This means that
results obtained with the GRAPE boards will have the full accuracy of
direct summation with 32 or 64 bit precision and argues that, in order
to increase the accuracy of the force calculation, one should increase
the number of particles in the realisation rather than consider a more
accurate potential solver, if, as is the case for GRAPE, the errors
are not systematic. 

Again using the $MASE$ values  
we find that the accuracy of the GRAPE tree code is comparable to that
of direct summation and 
that can be explained by the fact that contributions of nearby
particles are calculated in both cases in the same way.
As expected the force calculations are more accurate for smaller
tolerances or larger $n_{crit}$, but the effects are small. In order
to increase the accuracy it is thus more efficient to increase the
number of particles than to decrease the tolerance.

We also performed a number of simulations of the evolution of a
Plummer sphere in order to test the energy conservation, although the
latter does not depend only on the calculations on the GRAPE boards but also
on the time integration scheme and time step used. We find very good
energy conservation and discuss the influence of the softening on the
time step that should be used.

As a final test of the adequacy of GRAPE-3 boards to stellar dynamical
simulations we evolve the same initial conditions using different
hardware and software. These include the 64-bit precision direct
summation on GRAPE-4 boards and direct summation and tree code on 
GRAPE-3 boards. The simulation is the long term evolution of a barred
galaxy and we find that the pattern speeds as calculated by direct
summation on GRAPE-4 and on GRAPE-3 show excellent agreement. The
agreement is also very good with the tree code with small opening
angles. Using large opening angles is not as satisfactory.

One can thus conclude from all the above that GRAPE-3 boards are well
suited for simulations of galaxies or galaxy systems, both because
they have the necessary accuracy and because their high speed allows
the use of a large number of particles.

\section*{Acknowledgments} 
We would like to thank Eric Fady for his contributions to
the $MISE$ calculations and Jerry Sellwood for making available to us
his cartesian grid code.
We would also like to thank the
INSU/CNRS and the University of Aix-Marseille I for funds to develop
the computing facilities used for the calculations in this paper.
We gratefully acknowledge the use of joint CNRS - JSPS grants for
travel exchanges. 
The NEMO package was often used both in the generation of the initial
conditions and the analysis of the simulations and we are indebted to Peter
Teuben for his efforts to maintain it.

\appendix

\section[]{Asymptotic behaviour of the mean integrated square error}

The asymptotic behaviour of the $MISE$ or $MASE$  can be 
understood with a very simple argument as below. Let us first
discuss the ``bias'' part (Merritt 1996) of the error,
which is the error due to the finite softening in the limit of
$N\rightarrow \infty$, and then the effect of finite $N$,
or the ``variance'' part in the terms used by Merritt (1996).

In the following, we consider the softening in terms of the softening
kernel function $W(r/\epsilon)$. The exact force on a particle at
position ${\bf r}$ is given by
\begin{equation}
{\bf F}_{exact} = \int d{\bf r}' {\bf F}_{pp}({\bf r}, {\bf r}') 
\rho ({\bf r}').
\end{equation}
where ${\bf F}_{pp}= G \frac{{\bf r}'- {\bf r}}{|{\bf r}'- {\bf r}|^3}.$
The softened force is calculated as
\begin{equation}
{\bf F}_{soft} = \int d{\bf r}' \rho ({\bf r}') \int d{\bf x} W(x/\epsilon){\bf
F}_{pp}({\bf r}, {\bf r}' + {\bf x}).
\end{equation}
Since ${\bf F}_{pp}$ actually depends only on the relative position,
we can rewrite the above equation as
\begin{equation}
{\bf F}_{soft} = \int d{\bf r}'  \rho ({\bf r}') \int d{\bf x} W(x/\epsilon){\bf
F}_{pp}({\bf r}+ {\bf x}, {\bf r}' ).
\end{equation}
By changing the order of the space integration, we then find
\begin{equation}
{\bf F}_{soft} =  \int d{\bf x} W(x/\epsilon){\bf F}_{exact}.
\label{eq:bias_theory}
\end{equation}
Equation (\ref{eq:bias_theory}) has the same form as that used
in all definitions in smoothed particle hydrodynamics (SPH, e.g. 
Monaghan 1992), where $W$ is the smoothing, or interpolating kernel. 
Following the argument used in SPH (e.g. Hernquist \& Katz 1989), 
we can show that
\begin{equation}
{\bf F}_{soft}  = {\bf F}_{exact} + {\cal O}(\epsilon^2),
\label{eq:bias_asymptotic1}
\end{equation}
so that for the squared error we have
\begin{equation}
|{\bf F}_{soft} - {\bf F}_{exact}|^2 \propto  \epsilon^4.
\label{eq:bias_asymptotic2}
\end{equation}

For the ``variance'' part of a non-singular distribution of particles
we can write
\begin{equation}
|{\bf F}_{nbody} - {\bf F}_{soft}|^2 \propto (N\epsilon)^{-1}.
\label{eq:variance}
\end{equation}
Equation (\ref{eq:variance}) can be understood as follows: For a given
softening, if one changes $N$, the squared error should decrease as
$1/N$, since we can consider the N-body force as a Monte-Carlo
integration of the softened potential field. The argument for the
dependence on the softening is equally simple. 
The random error in the force is dominated by the variation
of the forces from particles with the distance of the order of the
softening parameter. The number of particles in that region is
proportional to $\epsilon^3$. On the other hand, the typical force
from one of these particles is proportional to $\epsilon^{-2}$. Thus,
the squared error is proportional to 
$\epsilon^{-1}$. Similar results, both for the ``variance'' and the
``bias'' have been found by Merritt \& Tremblay (1994)

To derive the ``optimum'' value of $\epsilon$, which
will minimise the sum of the bias and the variance, we write
\begin{equation}
MISE = c_1\epsilon^4 + \displaystyle{c_2 \over N\epsilon},
\end{equation}

\noindent
where $c_1$ and $c_2$ are constants. Thus, the optimal $\epsilon$ and 
$MISE$ are given by
\begin{equation}
\epsilon_{opt} \propto N^{-0.2},\\
\end{equation}
\begin{equation}MISE_{opt} \propto N^{-0.8}.
\end{equation}

\bsp

\label{lastpage}

\end{document}